\documentclass[fleqn]{article}
\usepackage[dvips]{graphicx}
\usepackage{emlines2,bezier,amscd,amsmath}
\usepackage{amssymb}

\title{Fall of an Elastic Bar in Central Gravitational Field: I. Newtonian Gravity }
\author{Sergey S. Kokarev\thanks{logos-center@mail.ru}}
\date{Regional Scientific Educational Center  "Logos"\, \\
Russia, 150000, Yaroslavl, Respublikanskaya 80, r.22\\
Research Institute of Hypercomplex Systems in Geometry and Physics,
(Moscow)
}

\begin{document}
\maketitle

\begin{abstract}
Within some reasonable approximations we calculate deformation of  an
elastic bar, falling on a source of central gravitational field.
We consider both elastic deformations and plastic flow together with  destruction of the bar.
Concrete calculations  for a number of materials are presented.
\end{abstract}


\section{Introduction}

The study of the extended deformable bodies motion in general
relativity (GR) is an important and technically challenging problem
with fundamental as well as practical applications. In order to be
consistent with principles of GR the solution requires a thorough
review of all the notions which are commonly used in nonrelativistic
continuous media physics. General principles lying at the foundation
of deformable bodies physics were actively debated when the
relativity theory first appeared  \cite{born,noet,laue}. Contemporary
understanding of the problem is based upon seminal works
\cite{surie,maugin,carter}. Different aspects of the elasticity
theory in GR were further developed in works \cite{MagKij} (gauge
formulation), \cite{Christodoulou} and \cite{beig,beig1,beig2}
(general theorems of  existence and uniqueness). The most
important and widely accepted topics in the field are: {\it material
(frame independent) description of elastic medium}, {\it
1+3-decomposition of space-time} and {\it Lagrangian formulation of gravitoelastic
equations}.

Note that more detailed treatment and application of these topics gave rise to slightly different
versions of the general approach, depending on author's conceptions; some of these
conceptions
are outside the scope of elasticity in GR  \cite{bel,samuelss,koks}.
Despite that these slightly different approaches to the problem reproduce the right nonrelativistic
limit, the predictions regarding experiments with extended elastic bodies based on them can diverge drastically.
That applies to the behavior of the bodies in strong gravitational fields (horizon of black hole) and even in
 weak fields without Newtonian limit (gravitational waves). So, the study of the behavior of deformable
  bodies in such exotic situations (from the point of view of Newtonian gravity) could serve
  for improvement and further development of elasticity theory in GR. At present both gravitational
  waves physics and black hole physics are theoretically well founded and actively developing topics.
  However, while the former has many intersections with elasticity in GR (first of all bar detectors and
  their interaction with wave perturbation in the space-time metric should be mentioned), incorporation of elasticity
  ideas into the black holes physics is much more modest. The majority of attempts in this or related fields dealt with
  the self-consistent problem of equilibrium of astronomical bodies that possess elastic properties
  (among the last works on the subject is already cited
\cite{samuelss}) and with tidal phenomena in celestial mechanics and astrophysics
\cite{ray,dobr,other1,other2}.

The present paper can be treated as a small step towards understanding what will be the influence of
a strong gravitational field  on test deformable bodies. Note that this problem is not the subject of
traditional elasticity GR since strong gravitational field causes tidal stresses which could lead to
plastic flow and mechanical destruction of any real body. We are going to illustrate our approach by
a relatively simple example of a one-dimensional continuous medium that is thin elastic bar, falling
towards the source of central gravitational field. We'll show that our treatment of the problem encounters
two difficulties which can be analyzed independently. The first difficulty originates from nonlinearity of
 the equations of motion and has nothing to do with relativistic aspects of the problem whatsoever.
 The second difficulty arises  from dealing with relativistic properties of both motion and space-time
 that have to taken into account in order to get a correct quantitative picture of deformations.

The aim of this part of the paper is to solve the first kind of difficulties within Newtonian gravity and to lay a
foundation for more precise and consistent consideration in the context of GR. We formulate an approximate "quasistatic fall"\,
approach which allows to solve nonlinear equations by the perturbations method, with smallness parameter
$L/r,$ where $L$ is the undeformed bar's length,
$r$ is instant distance from the bar to the source of gravitational field. The law of motion of point-like
particles will play the role of "null approximation". This nonrelativistic approach allows us to calculate the
deformational characteristics of bars up their destruction.
In section \ref{task} we derive the exact equation of motion (within Newtonian theory)
that describes the bar's behaviour including the deformation. In section \ref{approxx}
we discuss the exact formulation of the quasistatic fall condition. We find out that it is valid for bars made of solid materials.
In section \ref{appsol} we derive an approximate equation of motion in first order on $L/r$
and find solution to the derived equation that satisfies all necessary initial and boundary conditions.
Our approximate solution for elasticity reproduces the exact static solution (see
\cite{scheit})
after the transversal averaging and coarsening procedures.
Quantitative and qualitative deformational characteristics of the bar are calculated in section \ref{pict}.
In section \ref{formula} we derive slightly simpler formulae for calculation of definite
deformational characteristics: the time of the beginning of the plastic flow and the time of destruction
as well as the corresponding positions of the bar. Then, using the derived formulae, we find some "universal"\,
relations independent of the bar's initial state and present results acquired for some specified materials.
In the next part of the paper there will be presented relativistic analysis of the problem considering a deformable
body falling into Schwarzschild black hole.

\section{Statement of the problem} \label{task}

Let us consider radial fall of a thin probe bar from an initial position distanced from a massive
source of Newtonian central gravitational field. While in motion the bar stays oriented along the line of
force (in a radial direction).  Let us put the origin O of an one-dimensional coordinate system at
the force center and direct the coordinate axe OX along the bar's motion radial line.
Then kinematics of the bar can be described by a displacement field $x(t,\xi),$ defined on the points
$\xi$ of a remote unstrained bar (Lagrangian picture)\footnote{So,
we absolutely ignore
transversal elasticity of the bar.
Such simplifying is correct, since ratio
of absolute transversal $\Delta r$ and longitudinal
$\Delta l$ deformations satisfies the following nonequality:
\[
\frac{\Delta r}{\Delta l}\sim\mu\frac{R}{L}\ll1
\]
due to thinness of the bar ($R$ is its  characteristic transversal size) and
smallness of Poisson's coefficient for majority of solid materials. Exact results of \cite{scheit} support this
assumption: even for solid bar with $R/L=1/4$ ratio $\Delta r_{\text{max}}/\Delta l\sim 1/10$ near the Earth.}, and depending on time.
The initial and boundary conditions are as follows:
\begin{equation}\label{begin}
x(0,\xi)=r_0+\xi,\ \xi\in[-L/2;L/2];\quad \dot x(0,\xi)=-v_0=\text{const};\quad \sigma|_{\xi=0,L}=0.
\end{equation}
Here $r_0$ is the initial coordinate for the bar's  center of mass, $L$ is
the unstrained bar's length, $v_0$ is absolute value of initial
velocity, $\sigma=\sigma(t,\xi)=\sigma_{xx}$ is 1-dimensional "stress tensor"\,
(later to  be referred as  "stress") that depends on $\xi$ and $t.$
Here and below (if  not specified otherwise) the dot denotes differentiation
with respect to time coordinate  $t$ (or $\tau$ --- see below),  the accent denotes differentiation
with respect to space Lagrangian coordinate $\xi$.
To
be more precise, we define $\sigma$ as the following average value:
\[
\sigma=\frac{1}{S}\int\sigma^{\text{local}}_{xx}
\]
where integral is being taking over transversal section of the bar,
$S$ means  its area and
$\sigma^{\text{local}}_{xx}$ is true longitudinal  stress tensor
component, that depends on the points of the transversal section.

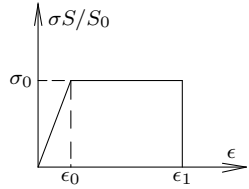
\begin{figure}[htb]
\centering \unitlength=0.50mm \special{em:linewidth 0.4pt}
\linethickness{0.4pt} \footnotesize \unitlength=0.70mm \special{em:linewidth
0.4pt} \linethickness{0.4pt}
\unitlength=1mm
\special{em:linewidth 0.4pt}
\linethickness{0.4pt}
\begin{picture}(30.67,25.00)
\emline{3.00}{25.00}{1}{3.00}{3.17}{2}
\emline{3.00}{3.17}{3}{30.67}{3.17}{4}
\emline{30.67}{3.17}{5}{27.50}{3.67}{6}
\emline{27.50}{2.50}{7}{30.67}{3.17}{8}
\emline{2.50}{22.00}{9}{3.00}{24.83}{10}
\emline{3.00}{24.83}{11}{3.50}{22.00}{12}
\emline{3.00}{3.17}{13}{7.33}{14.67}{14}
\emline{7.33}{14.67}{15}{22.17}{14.67}{16}
\emline{22.17}{14.67}{17}{22.17}{3.17}{18}
\emline{7.33}{14.67}{19}{7.33}{12.17}{20}
\emline{7.33}{10.67}{21}{7.33}{8.33}{22}
\emline{7.33}{6.50}{23}{7.33}{4.33}{24}
\emline{7.33}{3.50}{25}{7.33}{3.17}{26}
\emline{6.50}{14.67}{27}{4.83}{14.67}{28}
\emline{4.17}{14.67}{29}{3.00}{14.67}{30}
\put(4.33,22.83){\makebox(0,0)[lc]{$\sigma S/S_0$}}
\put(28.67,4.33){\makebox(0,0)[cb]{$\epsilon$}}
\put(2.33,14.67){\makebox(0,0)[rc]{$\sigma_0$}}
\put(7.33,2.50){\makebox(0,0)[ct]{$\epsilon_0$}}
\put(22.17,2.50){\makebox(0,0)[ct]{$\epsilon_1$}}
\end{picture}
\caption{\small Simplified stress diagram of bar's material, expressing results of
laboratory experiments with real bar.}\label{diag}
\end{figure}

Lets us consider small  element $dx$ of the bar at some fixed moment of time $t.$
Its ends have material coordinates $\xi$ and $\xi+d\xi.$
Inertial force, related to the element is expressed by the formula:
\[
-dm\, \ddot x=-\rho S\,\ddot x\,dx,
\]
where $\rho=\rho(t,\xi)$ is  the local volume mass density of the bar.
Gravitational force, acting
on the same element, has the form:
\[
-\frac{GM\,dm}{x^2}=-\frac{GM\rho S\,dx}{x^2},
\]
where $M$ is mass of origin, $G$ is the Newtonian constant.
Finally, elastic force, acting on the element looks like this:
\[
\frac{\partial(\sigma S)}{\partial x}\,dx.
\]
After we've added the three forces together and equated the resultant to zero, the following equation
of motion can be obtained:
\begin{equation}\label{eq1}
\ddot x-\frac{1}{\rho S}\frac{\partial (\sigma S)}{\partial x}=-\frac{GM}{x^2}.
\end{equation}
In order to exclude stress it is necessary to set some defining relation
\cite{truesdell} for the material of the bar. Said
relation normally connects stress $\sigma$ with relative deformation
$\epsilon\equiv x'-1.$ In this paper we'll use  piecewise linear defining relation, which
is commonly applied in the majority
of the model problems dealing with elasticity theory and  strength of materials \cite{rabotnov}.
Fig.\ref{diag} shows a  simplified effective diagram "strain-stress"\,
illustrating this defining relation.
In real laboratory experiments with bars the controlled external force is usually related to
the initial area $S_0$ of the transversal section before stress starts to build up. So,
such experiments allow only to determine dependency of $\sigma S/S_0$
on $\epsilon,$ which includes variation of the transversal section area
under deformation. The inclined part of the diagram  describes elastic deformations ({\it Hooke's law}),
for which $\sigma S/S_0=E\epsilon,$ where $E=\sigma_0/\epsilon_0$ is Young's modulus
for the material of the  bar,
$\epsilon_0$ is limit of elasticity, $\sigma_0$ is stress of
plastic flow. The latter is described by the horizontal part of the diagram,
where strain increases under practically  constant stress.
Parameter $\epsilon_1$ characterizes
ultimate relative deformation causing the destruction of the material.

Using the chosen  defining relation, the  term linked to stress
in  (\ref{eq1}) can be transformed (under elastic deformations)  as follows:
\[
-S_0\frac{\partial(E\epsilon)}{\partial x}=-S_0\frac{(E(x'-1))'}{\partial x/\partial \xi},
\]
where in the differentiation with respect to  $x$ we've turned to the Lagrange variable $\xi.$
Assuming that the bar is homogeneous, we have $E=\text{const}$ and
$\rho_0=\rho(0,\xi)=\text{const}.$ Using of the relation:
\[
\rho_0S_0=\rho(t,\xi)x'S,
\]
that expresses the law of mass conservation as applied  to  any element of the bar related to
fixed material particles, equation (\ref{eq1}) can be rewritten in the form:
\begin{equation}\label{eq2}
\Box x=-\frac{r_{\text{s}}}{2x^2},
\end{equation}
where $\Box\equiv\partial^2_\tau-\partial^2_\xi$ is
the two-dimensional D'Alembert operator, $\tau=c_0t,$
$c_0=\sqrt{E/\rho_0}$ is sound velocity in the material of the bar, $r_{\text{s}}=2GM/c_0^2$ is
sound analog of gravitational radius, which we'll call
the {\it sound radius}\footnote{In a difference with gravitational one, it depends on
elastic properties of bodies, falling on the gravitational center.}.

Equation (\ref{eq2}) is nonlinear wave equation, for which
standard methods of ma\-the\-ma\-ti\-cal physics are inapplicable.
It easy to find a family of particular soliton-like solutions, which is characterized by special
dependency on $\xi$ and $\tau$: $x=x(\zeta),\ \zeta=a\tau+b\xi,$ where
$a$ and $b$ are arbitrary constants,
satisfying the condition: $|a|\neq|b|.$ Substitution to (\ref{eq2})
leads to an ordinary differential equation:
\begin{equation}\label{traj}
x''+\frac{k}{x^2}=0,
\end{equation}
with $k=r_{\text{s}}/2(a^2-b^2),$ where the accent denotes differentiation with respect to $\zeta.$
However, its general solution:
\begin{equation}\label{gsol}
-\frac{1}{A}\sqrt{(Ax+2k)x}+
\frac{k}{A^{3/2}}\ln(k+Ax+\sqrt{Ax(Ax+2k)})=\zeta-\zeta_0,
\end{equation}
where $A$ and $\zeta_0$ are integration constants,
doesn't satisfy conditions (\ref{begin}),
no matter what constants or parameters are being chosen.
In general, our study will not involve
those stressed states of the bar that are associated with running waves.
We'll restrict ourselves to seeking  solutions to  the equation (\ref{eq2}) that  describe a smooth  (quasi-static)
variation of the bar's  stressed state from null strain (far from the gravitational center) up to
the point of destruction. Note, that for those parts of the bar, where elasticity limit is exceeded,
the equation (\ref{eq2}), according  to the  stress diagram in
Fig.\ref{diag},
will acquire a simpler form as that  of an ordinary differential equation similar to (\ref{traj}).

\section{Condition of quasi-static fall} \label{approxx}

On an obvious
assumption that the law of motion for material points roughly applies to the motion of extended small-sized bodies,
an approximate solution to the equation (\ref{eq2})
can be constructed. To be more precise,  we state that:
\begin{equation}\label{point}
\lim\limits_{L\to 0}\max\limits_{t\in[0;T]}|x(t)-x_0(t)|=0,
\end{equation}
where $L$ is characteristic size of the extended body, $x(t)$ is exact law of motion
for the body's center of mass, calculated  for time interval $[0;T]$ with regard to its extended
structure,
$x_0(t)$ is  law of motion for  point-like body with the same integral characteristic
(mass, charge etc.), calculated for the same time interval. The relation (\ref{point}) is, in fact,
one of the basic statements of both classical mechanics and relativity theory in implicit form.
(Precise formulations and important rigorous theorems see in
\cite{beig1}).

This assumption allows us to seek solution to the equation (\ref{eq2})  as follows:
\begin{equation}\label{quasi0}
x(\tau,\xi)=x_0(\tau)+\xi+\chi(\tau,\xi),
\end{equation}
where $x_0(\tau)+\xi$ describes the motions of the rigid bar (tidal forces to be disregarded), which center of mass initially is in the position
$x_0(0)=r_0,$ while correction $\chi(\tau,\xi)$
describes strain of the bar, subjected to in some set mode of motion $x_0(\tau).$
After we've calculated the strain  in a set mode of motion, if necessary, we can define the mode of fall more accurately and recalculate
the strain etc.

In order to establish the limits of validity of our approach, let us
estimate critical distance $r_{\text{c}},$
over which the   bar is destroyed.
Upon equating the characteristic tidal stress to ultimate stress, which is for majority of
solid materials three orders less then their Young's modulus,
our estimation appears as follows:
\[
r_{\text{c}}\sim10(r_{\text{g}}L^2)^{1/3}(c/c_0)^{2/3},
\]
where $r_{\text{g}}=2GM/c^2$ is the gravitational radius of the field's source.
Assuming that   $c/c_0\sim10^{5},$ we obtain a more convenient formula for calculating the
estimations:
\begin{equation}\label{cr}
r_{\text{c}}\sim 10^4(r_{\text{g}}L^2)^{1/3}.
\end{equation}
It follows from (\ref{cr}) that
for  gravitational centers whose  mass is  $M\leq10^3M_{\odot}$ and bars
of the
lengths $L\sim10^0$m
$r_{\text{c}}>r_{\text{g}}$ (for the Sun $r_{\text{g}}\sim1$km and $r_{\text{c}}\sim100$km).

The validity condition of the approach discussed  above is the "quasi-static fall"\, quality:
time $\Delta T$ for characteristic variation
of the tidal force must exceed by far the time  $L/c_0$
for the  longitudinal sound waves propagation  along the bar.
The  time $\Delta T$ can be roughly estimated  as $x_0/c_0\dot x_0.$
As the expression implies, the time  will be minimal at the moment of destruction.
The energy conservation law has the following dimensionless  form:
\begin{equation}\label{conserv}
\dot x_0^2=\frac{r_{\text{s}}}{x_0}+\frac{v_0^2}{c_0^2}-\frac{r_\text{s}}{r_0}
\end{equation}
The first term on the right at the moment of destruction  has the order $10^7$ (when $L\sim10^0$m,
$M\sim M_\odot$), while the second and third under nonrelativistic initial velocities
and large enough (compared to the gravitational radius of the source) initial distances
have considerably  less order.
Thus we  can disregard them at the moment of destruction
and  applying  (\ref{cr}) we get:
\begin{equation}\label{quasi}
\Delta T_{\text{min}}\sim10^{3/2}\frac{L}{c_0}>\frac{L}{c_0}.
\end{equation}
So, for bars made of solid materials (metals, steels and alloys, for
which $\sigma_1\sim10^{-3}E$ and $c_0\sim10^{-5}c$) the condition of
the fall's quasi-static quality is met automatically by application
of its  law of motion. For softer materials the
condition can be violated near the point of destruction. That implies
the necessity of studying  the shock waves and sound retarding impact on the problem.

\section{Approximate solution}\label{appsol}

After we have decomposed the gravitational force in the row (with respect to  $L/x_0$) near the   rigid bar's center of mass (its instant position)
$x_0(\tau)$, the righthand side of the equation (\ref{eq2}) can be rewritten in the following form:
\begin{equation}\label{decomp}
-\frac{r_{\text{s}}}{2(x_0+\xi)^2}=-\frac{r_{\text{s}}}{2x_0^2}+\frac{r_{\text{s}}\xi}{x_0^3}+o\left(\frac{L}{x_0}\right).
\end{equation}
After substituting it into  (\ref{eq2}) and taking (\ref{quasi0}) into account, we obtain
the
following approximate equation of motion:
\[
\ddot x_0+\ddot\chi-\chi''=-\frac{r_{\text{s}}}{2x_0^2}+\frac{r_{\text{s}}\xi}{x_0^3}.
\]
The first terms on the left and on the right
are cancelled as the  definition of $x_0(\tau)$ implicates.
The second term on the left is responsible for
{\it back reaction shown by  motion on the deformation of the bar}.
The term is of the order  $o(L^2/x_0^2)$ that can be proven later
when the approximate solution for $\chi$ is obtained, meaning that we can disregard it for the time being.
Then the equation that define strains,
takes the form:
\[
\chi''=-\frac{r_{\text{s}}\xi}{x_0^3}.
\]
Its general solution is:
\[
\chi=-\frac{r_{\text{s}}\xi^3}{6x_0^3}+C_1(\tau)\xi+C_2(\tau),
\]
where $C_1,C_2$ are arbitrary functional constants of integration.
Solution $x(\tau,\xi),$ that satisfies both initial (with accuracy $L^3/r_0^3$) and boundary conditions (\ref{begin}),
takes the form:
\begin{equation}\label{solution}
x(\tau,\xi)=x_0(\tau)+\xi(1+\frac{r_{\text{s}}}{2x_0^3}\left(\frac{L^2}{4}-\frac{\xi^2}{3}\right)).
\end{equation}

\section{Elastic and plastic deformations pictures of the  falling bar}\label{pict}

\begin{figure}[bht]
\centering
\includegraphics[width=0.4\textwidth]{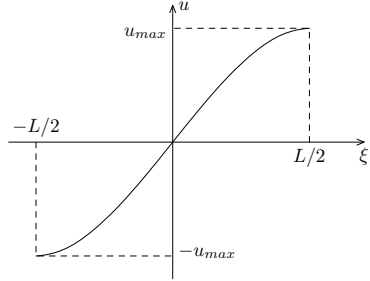}
\caption{\small Instant displacements diagram related to the center of the bar.
$u_{\text{max}}=r_{\text{s}} L^3/24x_0^3$.}\label{epdsp}
\end{figure}

Upon defining  the  displacements field  $u(\tau,\xi)$ by the formula: $x=x_0+\xi+u(\xi)$,
the solution  (\ref{solution}) results in expression:
\begin{equation}\label{displ}
u(\tau,\xi)=\frac{r_{\text{s}}\xi}{2x_0^3}\left(\frac{L^2}{4}-\frac{\xi^2}{3}\right).
\end{equation}
Fig.\ref{epdsp} shows the displacement diagram at some arbitrary (but preceding the plastic flow)
moment of time.

\hspace{3em}

\begin{figure}[htb]
\centering
\includegraphics[width=0.4\textwidth]{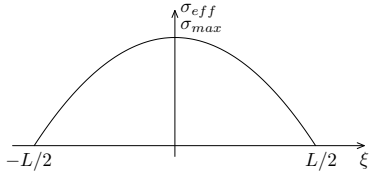}
\caption{\small Instant effective stress diagram before plastic flow.
$\sigma_{\text{max}}=Er_{\text{s}} L^2/8x_0^3$.}\label{estr}
\end{figure}

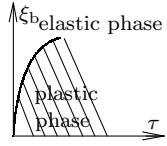
\begin{figure}[htb]
\centering \unitlength=0.50mm \special{em:linewidth 0.4pt}
\linethickness{0.4pt} \footnotesize \unitlength=0.70mm \special{em:linewidth
0.4pt} \linethickness{0.4pt}
\unitlength=0.70mm
\special{em:linewidth 0.4pt}
\linethickness{0.4pt}
\begin{picture}(30.51,27.16)
\emline{1.06}{23.50}{1}{1.84}{27.16}{2}
\emline{1.84}{27.16}{3}{2.51}{23.50}{4}
\emline{26.40}{2.61}{5}{30.40}{1.83}{6}
\emline{30.40}{1.83}{7}{26.40}{1.05}{8}
\put(2.95,24.94){\makebox(0,0)[lc]{$\xi_\text{b}$}}
\emline{27.01}{1.83}{9}{30.51}{1.83}{10}
\emline{29.84}{1.83}{11}{30.51}{1.83}{12}
\emline{1.84}{1.83}{13}{1.84}{27.00}{14}
\emline{1.83}{1.83}{15}{30.50}{1.83}{16}
\put(28.50,3.17){\makebox(0,0)[cb]{$\tau$}}
\bezier{100}(1.83,1.83)(2.83,18.33)(10.67,20.50)
\emline{2.83}{9.17}{17}{5.67}{1.83}{18}
\emline{3.67}{12.00}{19}{7.83}{1.83}{20}
\emline{10.17}{1.83}{21}{4.67}{14.67}{22}
\emline{12.50}{1.83}{23}{6.00}{16.67}{24}
\emline{15.00}{1.83}{25}{7.50}{18.67}{26}
\emline{17.33}{1.83}{27}{9.33}{20.00}{28}
\emline{19.67}{1.83}{29}{11.50}{20.33}{30}
\put(5.83,9.67){\makebox(0,0)[lc]{\text{plastic}}}
\put(6.00,4.50){\makebox(0,0)[lc]{\text{phase}}}
\put(6.67,23.00){\makebox(0,0)[lc]{\text{elastic phase}}}
\end{picture}
\caption{\small Boundary of elastic and plastic phases in falling bar as function of time.}\label{phase}
\end{figure}

The diagram during the fall is stretched  in vertical direction.
The instant effective stress diagram, expressed by the formula:
\begin{equation}\label{stress}
\sigma_{\text{eff}}\equiv\frac{\sigma S}{S_0}=\frac{r_{\text{s}} E}{2x_0^3}\left(\frac{L^2}{4}-\xi^2\right)
\end{equation}
is shown in Fig.\ref{estr}.
As the previous diagram, this one shows tendency to  vertical
stretching of the
same time law. At the moment $\tau^\ast,$
defined by relation:
\begin{equation}\label{time1}
x_0(\tau^\ast)=\left(\frac{r_{\text{s}} L^2}{8\epsilon_0}\right)^{1/3}
\end{equation}
stress   reaches
the first stress  limit $\sigma_0$ at the middle point of the bar and this marks
the nucleus of the plastic phase.
The size  $2\xi_{\text{b}}$ of the part of the bar engaged in the plastic phase increases in time symmetrically
in both directions from the center of the bar according to the expression:
\begin{equation}\label{plastic}
\xi_{\text{b}}(\tau)=\sqrt{\frac{L^2}{4}-\frac{2\sigma_0x_0^3(\tau)}{r_{\text{s}} E}}.
\end{equation}
This dependency is shown in Fig. \ref{phase}.

When the front of the plastic phase part approaches the ends of the bar,
its velocity $d\xi_{\text{b}}/dt$ asymptotically
tends to zero.
Any element of the bar in the plastic phase (i.e. situated inside the interval $[-\xi_\text{b}(\tau),\xi_{\text{b}}(\tau)]$
at the moment $\tau$)
moves free in gravitational field,
as the  stress diagram gives  $\sigma'=0$ for any point of plastic phase.

Since  the appearence of the plastic phase, matter elements situated  in the vicinity  of
the middle point of the bar, continue to suffer the stretch strain. However, they doesn't stretch
the same way the parts of elastic phase do (i.e. in accordance to the  solution (\ref{solution})),
but as particles
of noninteracting "dust packet". It is in the central point $\xi=0$ the relative plastic deformation will reach its  maximum $\epsilon_1$
at the moment $\tau^{\ast\ast},$
causing the  bar to break in two halves.

In order to calculate the relative deformation at the middle point of the bar at an arbitrary moment
of the plastic phase, it is worth noting that the element $d\xi$ of the bar containing middle point, will
by the beginning of the  plastic phase acquire the  length $ds=(\epsilon_0+1)\,d\xi$.
Its further  stretching can be described by the equation:
\[
ds'=\left.\frac{\partial \bar x_0(\tau,s)}{\partial s}\right|_{s=0}ds=\left.\frac{\partial
\bar x_0(\tau,s)}{\partial s}\right|_{s=0}(\epsilon_0+1)\,d\xi.
\]
Here $\bar x_0(\tau,s)$ means a family of solutions, that describe
free particles with coordinates $x_0(\tau^\ast)+s$
at the moment $\tau=\tau^\ast$ and velocity
$\dot x_0(\tau^\ast).$ Relative deformation with respect to initial
configuration is given by the expression:
\[
\epsilon'=\frac{ds'}{d\xi}-1=\left.\frac{\partial \bar x_0(\tau,s)}{\partial s}\right|_{s=0}(\epsilon_0+1)-1.
\]
Then equation for $\tau^{\ast\ast}$ takes the form:
\begin{equation}\label{destroy}
\left.\frac{\partial \bar x_0(\tau^{\ast\ast},s)}{\partial s}\right|_{s=0}=
\frac{\epsilon_1+1}{\epsilon_0+1}.
\end{equation}

Equations
(\ref{displ})-(\ref{destroy}) together with
the solution (\ref{gsol}) for free point-like particles give
a complete picture of deformations of the bar up to its destruction.

\section{Design formulae and  calculations}\label{formula}

Before we start deriving design  formulae for calculations we want to
stress that any more or less distinctive relative deformations of bar (say, about 0.01\%)
appears
only at the latest times, when the bar moves considerably close to
the gravitational center. So, for the  source of mass about of the Sun's and the bar
about 1m long we can expect by (\ref{stress}) the deformation
$\bar\epsilon=10^{-4}$ only at distance
of about $\bar r\sim 100$km from the gravitational center. The general formula is:
\[
\bar r\sim\left(\frac{r_{\text{s}}L^2}{\bar\epsilon}\right)^{1/3}
\]
It gives us practically the same result in the majority of
the other reasonable situations. So, the largest and most important part
of the bar's deformational history takes place under $x\lesssim\bar r.$
It implies, that
instead of the general solution (\ref{gsol}) we can use its simpler
asymptotic form under small $x,$ where the law of motion is parabolic:
\begin{equation}\label{parab}
x_0(\tau)=\left(\frac{9r_{\text{s}}}{4}\right)^{1/3}(\tau_0-\tau)^{2/3}=\left(\frac{9r_{\text{g}}c^2}{4}\right)^{1/3}(t_0-t)^{2/3},
\end{equation}
where $t_0=\tau_0/c_0$ means total time of particle's fall.
This is an unique value, which depends on the initial state of the falling object. It can be calculated
by means of the  exact formula (\ref{gsol}):
\begin{equation}\label{timef}
t_0=\frac{\sqrt{GMr_0/2}}{\varepsilon_0}\phi(\varepsilon_0 r_0/GM),
\end{equation}
where  $\varepsilon_0=E_0/m$  is initial specific energy of the falling object,
$r_0$ is initial distance from the gravitational center and the function
\[
\phi(z)=\sqrt{z+1}-\frac{1}{2\sqrt z}\ln(1+2z+2\sqrt{z(1+z)}).
\]
For $r_0\sim1$ a.u.,  $M\sim M_\odot$ and  $v=50$km/s the formula (\ref{timef}) gives
us $t_0\sim 25$ days.
The exact validity criterion  for the   parabolic law of motion near the source is as follows:
\[
\frac{\varepsilon_0}{c_0^2}\ll\left(\frac{r_\text{s}}{L}\right)^{2/3}\bar\epsilon^{1/3}.
\]
It is satisfied when  the initial velocities are nonrelativistic
and initial distances exceed $\bar r$ by far.

By using (\ref{traj}) and general formulae (\ref{displ})-(\ref{destroy})
of the previous section, it is easy to derive the following
convenient design  expressions, that describe
all characteristic events  of the bar's deformational history:
\begin{equation}\label{times}
\tau^\ast=\tau_0-\frac{L}{3\sqrt{2\epsilon_0}};\quad
\xi_\text{b}(\tau)=\sqrt{\frac{L^2}{4}-\frac{9\epsilon_0}{2}(\tau_0-\tau)^2}.
\end{equation}
Derivation of (\ref{destroy}) is slightly more complicated.
Decomposition of  (\ref{gsol}) in the vicinity of the parabolic law has the following form:
\begin{equation}\label{asparab}
C-\tau=\int\frac{dx}{\sqrt{A+r_\text{s}/x}}=\frac{2x^{3/2}}{3\sqrt{r_{\text{s}}}}-\frac{Ax^{5/2}}{5r_{\text{s}}^{3/2}}+o(A).
\end{equation}
Here $C$ and  $A$ are integration constants determined
by initial conditions at the moment of the appearance of the plastic phase:
\begin{equation}\label{boundary}
x(\tau^\ast)=x_0(\tau^\ast)+s;\quad
\dot x(\tau^\ast)=\dot x_0(\tau^\ast).
\end{equation}
Upon substituting  (\ref{asparab}) in (\ref{boundary}) and solving equations
with respect to $C$ and $A$
(in the first order with respect to  $s$)  we obtain  the following
implicit equation that  defines the law of motion in $s$-vicinity of the center
$\xi=0$ after
the moment of the appearance of the plastic phase:
\begin{equation}\label{asparsol}
\frac{2x^{3/2}}{3\sqrt{r_{\text{s}}}}-\frac{sx^{5/2}}{5\sqrt{r_{\text{s}}}x_0^{\ast2}}-\frac{4}{5}s\sqrt{\frac{x_0^\ast}{r_\text{s}}}=
\tau_0-\tau.
\end{equation}
Here $x_0^\ast\equiv x_0(\tau^\ast).$
After differentiating of the expression with respect to  $s$ under fixed  $\tau$ and assuming  $s=0$,
we obtain:
\[
\left.\frac{dx}{ds}\right|_{s=0}=\frac{4}{5}\sqrt{\frac{x_0^\ast}{x}}+\frac{1}{5}\frac{x^2}{x_0^{\ast2}}.
\]
Substituting in this expression $x(\tau)=x_0(\tau)$, we finally get:
\begin{equation}\label{diff}
\left.\frac{dx}{ds}\right|_{s=0}=\frac{4}{5}\sqrt{\frac{x_0^\ast}{x_0}}+\frac{1}{5}\frac{x_0^2}{x_0^{\ast2}}.
\end{equation}
Destruction conditions (\ref{destroy}) and (\ref{diff})  allow us to
come up with the following formula for the time of destruction:
\begin{equation}\label{destroyt}
\tau^{\ast\ast}=\tau_0-\frac{L\eta^3}{3\sqrt{2\epsilon_0}},
\end{equation}
where $\eta$ is  root of equation:
\[
\frac{4}{\eta}+\eta^4=\frac{5(\epsilon_1+1)}{\epsilon_0+1}.
\]

The dependency of all obtained formulae (\ref{times})-(\ref{destroyt})
on initial conditions is based on the  parameter $\tau_0$ only.
Some of consequences of these formulae are perfectly independent
of the initial conditions.
For example, full time of plastic phase existence:
\begin{equation}\label{timepl}
\Delta\tau\equiv\tau^{\ast\ast}-\tau^{\ast}=\frac{L}{3\sqrt{2\epsilon_0}}(1-\eta^3)
\end{equation}
and the size of the plastic phase at the moment of destruction:
\[
2\xi_{\text{b}}(\tau^{\ast\ast})=2\xi_{\text{b\,max}}=L\sqrt{1-\eta^6}
\]
depend only on the elastic constants for the material of the bar.
Also we have the following universal ratio of coordinates
$x^{\ast\ast}\equiv x(\tau^{\ast\ast})$ and
$x^\ast\equiv x(\tau^\ast)$:
\begin{equation}\label{conn}
x^{\ast\ast}=\eta^2x^\ast.
\end{equation}

Table 1 shows results of calculations made for the  bars of fixed length and of different
solid materials.
\bigskip

{\small
\hspace{-3em}\begin{tabular}{|l|c|c|c|c|c|c|c|}
\hline
Material&$\epsilon_0,10^{-2}\%$&$\epsilon_1,\%$&$x^\ast/r_{\text{g}}$&$\eta$&$x^{\ast\ast}/r_{\text{g}}$&$2\xi_{\text{b\,max}}/L$&$\Delta\tau/(\tau_0-\tau^\ast)$\\
\hline
Aluminium&3.1 &45&53 &0.56&17 &0.984&0.82\\ \hline
Iron  &8.5&50 &40&0.54&12 &0.987&0.84\\ \hline
Gold& 5.0&40&86 &0.58&29 &0.981&0.80\\ \hline
Copper&5.8 &60&54 &0.50&13.5 &0.992&0.88\\ \hline
Platinum&4.4&45&74 &0.56&23&0.984&0.82\\ \hline
Lead&3.1 &50&142 &0.54&41 &0.987&0.84\\ \hline
Silver&3.2 &45&82 &0.56&26&0.984&0.82\\ \hline
Titanium&9.1&70&39&0.47&9&0.996&0.90\\ \hline
Steel 15GS&17.5&18&31&0.78&19&0.880&0.53\\
(small strength)&&&&&&&\\ \hline
Steel N18K9M5T&97&8&17&0.90&14&0.684&0.27\\
(of high strength)&&&&&&&\\ \hline
Steel 30H13&75&6&19&0.92&16&0.627&0.22\\
(martensite)&&&&&&&\\ \hline
Alumalloy&49&10.5&31&0.86&23&0.772&0.36\\
(Al-Cu-Mg)D19T&&&&&&&\\ \hline
Titanalloy&108&4&20&0.95&18&0.514&0.14\\
(Ti-Al-V-Cr)TS6&&&&&&&\\ \hline
\end{tabular}

\medskip

Table 1. Results of approximate calculations by formulae (\ref{times}) and (\ref{destroyt})
for central body with $M=2\cdot10^{30}$kg and bars of length $L=1$m made of
different solid materials. Limit of elasticity $\epsilon_0$ calculated
by the formula: $\epsilon_0=\sigma_{0.2}/E,$
where $\sigma_{0.2}$ is (effective) stress which causes
residual relative deformation $0.2\%.$ Ultimate deformation $\epsilon_1=\epsilon_0+\delta,$
where $\delta$ is relative residual stretching after the destruction of the bar.
Steels labels corresponds to Russian standards \cite{sprav}.
}

\bigskip

From (\ref{time1})  and  (\ref{conn}) it follows that
immediately  after destruction the  plastic phase disappears
in both halves under the condition $\eta>1/2^{1/3}\approx0.794$.
Fifth column of the Table 1 shows that for all listed materials except high alloy steels,
the plastic phase will reappear
in future generations of the bar's fragments.
Inside of the bars  made of  high alloy steel, plastic phase appears,
increases, disappears
and after the fission reappears again in fragments some time later.
According to the first formula of (\ref{times}), bars with lengths of about $10^0$cm or less begins to flow
near the gravitational radius of the central body, which one should consider already a black hole.
More precise  and consequent calculations in
the context of GR will be presented  in next part of the paper.

\section{Conclusion}

We have considered intermediate nonrelativistic problem concerning
the fall of an elastic bar on massive source
of central gravitational field. Within the quasi-static fall approach
formulae (\ref{times}) and (\ref{destroyt})
we've been able to acquire reasonable and in some sense exhaustive description of the bar's deformational history up to  its
destruction. Note that our simple estimations
have shown that for
bars made of solid materials
like solid metals, steels and alloys the  quasistatic fall condition is satisfied automatically. Assumptions used for derivation of
(\ref{quasi}) from (\ref{conserv}) concern general nonrelativistic approximations of the problem.

Results of calculations, presented in Table 1 show that bars with length about 1m falling
on source with mass about $M_\odot$ are being destroyed far from the gravitational radius
(closest (of the  materials listed in Table 1) to it comes the titanium bar  --- about
10$r_\text{g}$).
Note that the time $\Delta\tau$ of the most important part of  the bar's deformational
history  is  about $1$ms in absolute units. After the first destruction each fragment of the bar gets divided again etc.
However, quantitative treatment of the behaviour of the generations of the bar's fragments   becomes more complicated,
because of non-elastic hysteresis and strain hardening phenomena,
that take place in deformed
fragments.
Taking
average values  $\bar\eta\approx0.67$ and $\bar x^\ast\approx53r_{\text{g}}$
from the Table 1,
and using the universal formula (\ref{conn}) we obtain:
\[
N\sim\log_{\bar\eta}\frac{1}{\sqrt{\bar x^\ast/r_{\text{g}}}}\approx
5
\]
--- a  rough estimation of number of the consequent fragments continuing dividing  until they reach
the gravitational radius.
Near the gravitational radius there will be required a cardinal reconsideration of our formulae within the context of GR.



 It is interesting,
that the  bar, falling in central gravitational field is, in some sense,
a perfect experimental device for obtaining
the exact stress diagrams. In  standard laboratory experiments
with bars initial homogeneous stressed state of the bar near ultimate stress  becomes complex nonhomogeneous one with
spontaneously appearing flowing neck. Unlike these, inside falling bar  we have the tidal gravitational stretching that reaches its maximum at
the center of the bar at any moment of time.
Due to  the volume rather than surface character of tidal forces,
the stress diagram obtained by experiments with falling bars, would be
independent of  the end's stressing and
Saint-Venant principle.

\end{document}